\documentclass[11pt]{article}
\usepackage{hyperref}
\usepackage{authblk}
\pdfoutput=1
\usepackage{hyperref}
\pdfoutput=1
\begin{document}
\title{Clustering of inertial cloud droplets in isotropic turbulence}
\author[1]{Peter J. Ireland}
\author[2]{John Clyne}
\author[2]{Perry Domingo}
\author[2]{Tim Scheitlin}
\author[1]{Lance R. Collins}

\affil[1]{Cornell University, Ithaca, NY, USA}
\affil[2]{U. S. National Center for Atmospheric Research, Boulder, CO, USA}
\maketitle
%% The abstract (in this file, and that submitted as text to arXiv) should
%% include the exact phrase
%% "fluid dynamics video" or "fluid dynamics videos"
\begin{abstract}
This article accompanies the submission of a fluid dynamics video (entry 102248)
of inertial cloud droplet clustering in isotropic turbulence
for the Gallery of Fluid Motion of the 66th Annual Meeting of the American Physical Society, 
Division of Fluid Dynamics.
\end{abstract}
% main text
\section{Numerical simulation}
We performed a direct numerical simulation of statistically stationary homogeneous isotropic turbulence 
on over 8 billion grid points, using 16,384 processors on the NCAR Yellowstone supercomputer [1].
The Taylor microscale Reynolds number for the flow is about 600, the highest Reynolds number ever
simulated for inertial-particle-laden isotropic turbulence. The domain is a cube of length $3500 \eta$,
and the flow evolves for a period of about $100 \tau_\eta$. ($\eta$ and $\tau_\eta$ are the Kolmogorov
length and time scales of the turbulence, respectively.) A subregion of the cube measuring approximately
$2000 \eta \times 2000 \eta \times 150 \eta$ is shown.

The flow was seeded with about 3 billion inertial (i.e., denser than the carrier fluid) cloud droplets.
We show a subset of these droplets in the video for two sizes: $St=1$ (corresponding to intermediate-sized
droplets $\sim 100$ $\mu$m in diameter), and $St=30$ (corresponding to rain-sized drops $\sim 0.5$ mm
in diameter).  The Stokes number $St$ is a non-dimensional measure of particle inertia.
The particle concentration field (isocontours) is visualized in blue,
and regions of high fluid vorticity (direct volume rendering) are visualized in yellow.
We notice that the intermediate-sized droplets ($St=1$) are centrifuged
out of regions of high vorticity, as originally suggested in [2].
As a result, they form small, dense clusters. The typical length of these
clusters is $10 \eta$ [3], or about $1/200^\mathrm{th}$ 
of the width of the subregion shown.
Large, rain-sized drops ($St=30$) are less responsive to the underlying vorticity field
and instead form much larger, less dense clusters.  The typical length of these
clusters is about $200 \eta$, or about $1/10^\mathrm{th}$ of the
width of the subregion shown.

We quantify the degree of particle clustering through the radial
distribution function (RDF). The RDF is the ratio of particle pairs at a given separation
to that of a uniform distribution.  We notice that $St=1$ particles
are tightly clustered, with clustering over 20 times that of a uniform
distribution at the smallest separations. The RDF rapidly drops off
at larger separations, suggesting that the particle clusters have small length scales.
$St=30$ particles, on the other hand, are loosely clustered 
(clustering is about twice that of a uniform distribution at the
smallest separations), but this clustering persists to larger scales, 
as is evident in the visualizations.

\section{Physical significance}
Standard microphysical cloud models are unable to predict the rapid onset of precipitation
in warm cumulus clouds [4]. The accelerated rate of precipitation has been linked to
turbulence-induced collisions within clouds [5], which are enhanced by clustering
due to the inertia of the droplets [6]. Our visualizations provide clear verification of this clustering
at the highest Reynolds number simulated to date. A detailed examination of the
physical processes is in preparation [7].

\section{Acknowledgements}
This work was supported by the National Science Foundation grant CBET-0967349 and
through a Graduate Research Fellowship to PJI. We would also like to acknowledge
high-performance computing support from Yellowstone  (ark:/85065/d7wd3xhc) provided by
NCAR's Computational and Information Systems Laboratory, sponsored by the National Science Foundation.
\section{References}
[1] Computational and Information Systems Laboratory. 2012. 
Yellowstone: IBM iDataPlex System (Climate Simulation Laboratory). 
Boulder, CO: National Center for Atmospheric Research. http://n2t.net/ark:/85065/d7wd3xhc.\\~
[2] {\sc M. R. Maxey.} 1987 The motion of small spherical particles in a celluar flow field. 
\emph{Phys. Fluids} {\bf 30}, 1915--1928.\\~
[3] {\sc A. Aliseda, A. Cartellier, F. Hainaux, and J. C. Lasheras.} 2002 
Effect of preferential concentration on the settling velocity of heavy particles in homogeneous isotropic turbulence.
\emph{J. Fluid Mech.} {\bf 468}, 77--105.\\~
[4] {\sc K. V. Beard and H. T. Ochs III.} 1993
Warm-rain initiation: An overview of the microphysical mechanisms.
\emph{J. Appl. Meteo.} {\bf 32}, 608--625.\\~
[5] {\sc R. A. Shaw.} 2003
Particle-turbulence induced interactions in atmospheric clouds.
\emph{Annu. Rev. Fluid Mech.} {\bf 35}, 183--227.\\~
[6] {\sc S. Sundaram and L. R. Collins.} 1997
Collision statistics in an isotropic, particle-laden turbulent suspension I.
Direct numerical simulations.
\emph{J. Fluid Mech.} {\bf 335}, 75--109.\\~
[7] {\sc P. J. Ireland, A. D. Bragg, and L. R. Collins.} 2013
Analyzing the effect of Reynolds number on inertial particle
dynamics in isotropic turbulence through direct numerical simulation.
In preparation.

\end{document}